\documentclass[nofootinbib,prd,twocolumn,showpacs,showkeys,preprintnumbers]{revtex4-1}
\usepackage{hyperref,amssymb,amsmath,mathrsfs,bm,graphicx}
\begin{document}

\title {Physical infeasibility of geodesic dissipative dust as a source of gravitational radiation}
\author{L. Herrera}
\email{lherrera@usal.es}
\affiliation{Escuela de F\'\i sica, Facultad de Ciencias, Universidad Central de Venezuela, Caracas, Venezuela and Instituto Universitario de F\'isica
Fundamental y Matem\'aticas, Universidad de Salamanca, Salamanca, Spain}
\author{A. Di Prisco}
\email{alicia.diprisco@ciens.ucv.ve}
\affiliation{Escuela de F\'\i sica, Facultad de Ciencias, Universidad Central de Venezuela, Caracas, Venezuela}
\author{J. Ospino}
\email{j.ospino@usal.es}
\affiliation{Departamento de Matem\'atica Aplicada and Instituto Universitario de F\'isica
Fundamental y Matem\'aticas, Universidad de Salamanca, Salamanca, Spain}
\begin{abstract}
Using a framework based on the $1+3$ formalism, we show that a  source represented by a geodesic, dissipative, rotational dust, endowed with axial and reflection symmetry, violates regularity conditions at the center of the fluid distribution, unless the dissipative flux  vanishes. In this latter case the vorticity also must vanish, and  the resulting spacetime is Friedman--Robertson--Walker (FRW). Therefore it does not produce gravitational radiation.
\end{abstract}
\maketitle

\section{Introduction}
In a recent paper \cite{1}, the $1+3$ formalism \cite{21cil, n1, 22cil, nin} has been used to develop a general framework for studying axially symmetric dissipative fluids. More recently, the above mentioned framework has been  applied to study the perfect and geodesic, fluid case \cite{2}, without imposing from the beginning the shear--free condition. However, as a result of this study, it appears that all possible models, compatible with the assumed line element,  are in fact  shear--free, and according to the result obtained in \cite{3}, they are Friedman--Robertson--Walker (FRW), i.e. they do not produce gravitational radiation.

We recall that  we define a state of intrinsic
gravitational radiation (at any given point), to be one in
which  the super-Poynting vector does not vanish for any  unit timelike vector   \cite{bel, agp, Mac}. Then since the vanishing of the magnetic part of the Weyl tensor implies the vanishing of the super-Poynting vector, it is clear that  FRW does not produce gravitational radiation. It is also  worth recalling that the tight link between the super-Poynting vector and the existence of a state of radiation, is firmly supported by the relationship between the former and the Bondi news function \cite{7, 8} (see \cite{why} for a discussion on this point).

The fact that a non--dissipative source, does not radiate gravitational waves, becomes intelligible if we recall that radiation is an irreversible process. This  emerges at once  if  absorption is taken into account and/or Sommerfeld type conditions, which eliminate inward traveling waves, are imposed \cite{7}. Accordingly,   an entropy generator factor should be present in the description of the source,  a factor which of course is absent in a perfect fluid, and even more so in a collisionless dust. In other words, we should expect the irreversibility of the  process of emission of gravitational waves, to be represented in the equation of state through an entropy increasing  (dissipative) factor.

From the comments above, it becomes clear that  the simplest fluid distribution which we might believe to be  compatible with a gravitational radiation, is a dissipative dust under the geodesic condition. Such a case will be analyzed in this work.

We shall consider separately the two  possible subcases, namely: the fluid distribution is assumed, from the beginning, to be vorticity--free, or not. 

In the former case, it is shown that the vanishing vorticity implies the vanishing of the heat flux vector, and therefore, as shown in \cite{2}, the resulting spacetime is FRW.

In the latter case, it is shown that the enforcement of the regularity conditions  at the center, implies the vanishing of the dissipative flux, leading also to a FRW spacetime.

Thus  all possible models, sourced by a dissipative  geodesic dust fluid, belonging to the family of the line element considered here, do not radiate gravitational waves during their evolution, unless regularity conditions at the center of the distribution are relaxed.

We shall discuss about this result and its relationship with the peculiar evolution of the vorticity under the geodesic condition.

This work heavily relies on the general framework developed in \cite{1}, keeping the same notation, and just reducing the general equations to the particular case considered here.  These will be presented in an Appendix.

\section{The dissipative, geodesic, dust}
Let us consider  axially and reflection symmetric geodesic dust distributions (not necessarily bounded). For such a system the  line element  in ``Weyl spherical coordinates'', is assumed  as in \cite{2}:
\begin{equation}
ds^2=-dt^2+B^2\left  (dr^2+r^2d\theta ^2\right)+2\tilde G(r,\theta)dt d\theta+C^2d \phi ^2,\label{metric2}
\end{equation}
where  $B$ and $C$,  are positive functions of $t$, $r$ and $\theta$, and $\tilde G$ depends only on $r, \theta$. We number the coordinates $x^0=t, x^1=r, x^2= \theta, x^3=\phi$.

The energy momentum tensor in the ``canonical'' form reads:
\begin{eqnarray}
{T}_{\alpha\beta}&=& \mu V_\alpha V_\beta+q _{\alpha} V_{\beta}+q _{\beta} V_{\alpha} ,
\label{6bis}
\end{eqnarray}
where as usual, $\mu, q_{\alpha}$ and  $V_\beta$ denote the energy density, the heat flux vector and  the four velocity, respectively.

The heat flux vector can be decomposed, in terms of two scalar functions $q_I, q_{II}$, as (see \cite{1} for details)
\begin{equation}
q_\mu=q_IK_\mu+q_{II} L_\mu,
\label{qn1}
\end{equation}
\noindent with vectors $\bold K$ and $\bold L$  having components:
\begin{equation}
K_\alpha=(0, B, 0, 0);\qquad L_\alpha=(0, 0, \sqrt{B^2r^2+\tilde G^2}, 0).\label{vec}
\end{equation}
Next, the shear tensor
\begin{equation}
\sigma_{\alpha \beta}= V_{(\alpha;\beta)}+a_{(\alpha}
V_{\beta)}-\frac{1}{3}\Theta h_{\alpha \beta}. \label{acc}
\end{equation}
where
\begin{equation}
h^\alpha_{\beta}=\delta ^\alpha_{\beta}+V^\alpha V_{\beta},
\label{vel5}
\end{equation}
 may be  defined through two scalar functions, as:

\begin{eqnarray}
\sigma _{\alpha \beta}=\frac{1}{3}(2\sigma _I+\sigma_{II}) (K_\alpha
K_\beta-\frac{1}{3}h_{\alpha \beta})\nonumber \\+\frac{1}{3}(2\sigma _{II}+\sigma_I) (L_\alpha
L_\beta-\frac{1}{3}h_{\alpha \beta}).\label{sigmaT}
\end{eqnarray}
The above scalars may be written in terms of the metric functions and their derivatives as (see \cite{1}):
\begin{equation}
\sigma _I=\frac{B^2r^2+2\tilde G^2}{B^2r^2+\tilde G^2}\frac{\dot B}{B}-\frac{\dot C}{C},\label{sn}
\end{equation}
\begin{equation}
\sigma_I-\sigma_{II}=\frac{3\tilde G^2}{B^2r^2+\tilde G^2}\frac{\dot B}{B}.\label{sigmasa}
\end{equation}

For the other kinematical variables (the expansion  and the vorticity) we have:

\begin{equation}
\Theta=\frac{2B^2r^2+\tilde G^2}{B^2r^2+\tilde G^2}\frac{\dot B}{B}+\frac{\dot C}{C},\label{Thetaa}
\end{equation}
whereas the vorticity   may be described, either by the vorticity vector  $\omega^\alpha$, or the vorticity tensor $\Omega
^{\beta\mu}$, defined as:
\begin{equation}
\omega_\alpha=\frac{1}{2}\,\eta_{\alpha\beta\mu\nu}\,V^{\beta;\mu}\,V^\nu=\frac{1}{2}\,\eta_{\alpha\beta\mu\nu}\,\Omega
^{\beta\mu}\,V^\nu,\label{vomega}
\end{equation}
where $\Omega_{\alpha\beta}=V_{[\alpha;\beta]}+a_{[\alpha}
V_{\beta]}$, and $\eta_{\alpha\beta\mu\nu}$ denotes  the Levi-Civita tensor.

 We find a single component different from zero,  producing:

\begin{equation}
\Omega_{\alpha\beta}=\Omega (L_\alpha K_\beta -L_\beta
K_{\alpha}),\label{omegaT}
\end{equation}
and
\begin{equation}
\omega _\alpha =-\Omega S_\alpha.
\end{equation}
with the scalar function $\Omega$ given by
\begin{equation}
\Omega=\frac{\tilde G^\prime}{2 B\sqrt{B^2r^2+\tilde G^2}}.
\label{no}
\end{equation}

Observe that from (\ref{no}) and regularity conditions at the centre, it follows that: $G=0\Leftrightarrow \Omega=0$.

Regularity conditions at the center, necessary to ensure elementary flatness in the vicinity of the axis of symmetry \cite{1n, 2n, 3n}, require that as $r\approx 0$
\begin{equation}
\Omega=\sum_{n \geq1}\Omega^{(n)}(t,\theta) r^{n},
\label{sum1}
\end{equation}
implying, because of (\ref{no}), that in the neighborhood of the center
\begin{equation}
\tilde G=\sum_{n\geq 3}\tilde G^{(n)}(\theta) r^{n},
\label{sum1b}
\end{equation}
This last result in turn implies that as $r$ aproaches $0$, 

\begin{equation}
 \sigma_I- \sigma_{II}=\sum_{n\geq 4}\left[\sigma_I^{(n)}(t,\theta)-\sigma_{II}^{(n)}(t,\theta) \right]r^{n}.
\label{ref1}
\end{equation}

Now, from the elementary flatness condition, we may write, as $r\rightarrow 0$, 

\begin{equation}
C\approx r\gamma(t,\theta), 
\label{nuev1}
\end{equation}
implying
\begin{equation}
C^\prime \approx \gamma(t,\theta), \;\; C_{,\theta}\approx r\gamma_{,\theta},
\label{nuev3}
\end{equation}
where $\gamma(t,\theta)$ is an arbitrary function of its arguments.

We shall treat separately, the two possible cases, with and without vorticity.

Let us  first consider the restricted case $\tilde G=0\Leftrightarrow \Omega=0$.
\section{The vorticity free case}
In order to prove the main result of this work, i.e., that a  source represented by a geodesic, dissipative, dust, endowed with axially and reflection symmetry, is just Friedman--Robertson--Walker (FRW), we shall first consider the case when the dust is irrotational.

Thus, if we assume the vorticity to vanish, then, from, (\ref{sn}, \ref{sigmasa}, \ref{Thetaa}) we get
\begin{equation} 
\Theta=2\frac{\dot B}{B}+\frac{\dot C}{C},\;\sigma_I=\sigma_{II}=\tilde\sigma=\frac{\dot B}{B}-\frac{\dot C}{C}\label{sigmas'},
\end{equation}
implying:
\begin{equation}
\Theta=2\tilde \sigma+\frac{3\dot C}{C}=\frac{3\dot B}{B}-\tilde \sigma.
\label{nthet}
\end{equation}
In this case (\ref{ec2}) produces the following two equations
\begin{equation}
\dot q_I+\frac{q_I}{3}(4 \Theta+\tilde \sigma)=0,
\label{mot1}
\end{equation}
\begin{equation}
\dot q_{II}+\frac{q_{II}}{3}(4 \Theta+\tilde \sigma)=0.
\label{mot2}
\end{equation}
Using (\ref{sigmas'}) in (\ref{mot1}, \ref{mot2}), we obtain at once
\begin{equation}
q_IB^3rC=\alpha_1(r,\theta);\qquad q_{II}B^3C=\alpha_2(r,\theta),
\label{vf1}
\end{equation}
implying
\begin{equation}
q_I=q(r,\theta)q_{II},
\label{vfn}
\end{equation}
where $\alpha_1(r,\theta), \alpha_2(r,\theta)$ are arbitrary functions of their arguments, and 
\begin{equation}
q(r,\theta)=\frac{\alpha_1}{\alpha_2 r}.
\label{vfn1}
\end{equation}

Next from (\ref{45Y}) in the vorticity free case, we obtain:
\begin{equation}
(q_IBrC)^\prime+ (q_{II}BC)_{,\theta}=0.
\label{vf2}
\end{equation}
The combination of (\ref{vf1}) with (\ref{vf2}) produces
\begin{equation}
\frac{B^\prime}{B}\alpha_1+\frac{B_{,\theta}}{B}\alpha_2=\frac{\alpha_1^\prime+\alpha_{2,\theta}}{2},
\label{vf3}
\end{equation}
which implies
\begin{equation}
B=T(t)\tilde B(r,\theta),
\label{vf4}
\end{equation}
where $T$ and $\tilde B$ are arbitrary functions of their arguments.

Next, (\ref{ecc5Y}, \ref{ecc6Y}, \ref{ecc8Y}, \ref{ecc7Y}), read, in this case, respectively:
\begin{equation}
\frac{1}{3B}\left(2\Theta^\prime-\tilde \sigma^\prime-\tilde \sigma\frac{3C^\prime}{C}\right)=8\pi q_I,
\label{vf5}
\end{equation}
\begin{equation}
\frac{1}{3Br}\left(2\Theta_{,\theta}-\tilde \sigma_{,\theta}-\tilde \sigma\frac{3C_{,\theta}}{C}\right)=8\pi q_{II},
\label{vf6}
\end{equation}
\begin{equation}
H_1=-\frac{\tilde \sigma}{2Br}\left(\frac{\tilde \sigma_{,\theta}}{\tilde \sigma}+\frac{C_{,\theta}}{C}\right)=-\frac{(\tilde \sigma C)_{,\theta}}{2BrC},
\label{vf7}
\end{equation}

\begin{equation}
H_2=\frac{\tilde \sigma}{2B}\left(\frac{\tilde \sigma^\prime}{\tilde \sigma}+\frac{C^\prime}{C}\right)=\frac{(\tilde \sigma C)^\prime}{2BC}.
\label{vf8}
\end{equation}

Combining (\ref{vf7}) with (\ref{vf8}), we obtain 
\begin{equation}
(H_1BrC)^\prime+(H_2BC)_{,\theta}=0,
\label{vf9}
\end{equation}
where $H_1, H_2$ are the two scalars defining the  magnetic part of the Weyl tensor (see Eq.(37) in \cite{1}).

From (\ref{vf5})  and (\ref{vf6}), it follows that
\begin{equation}
2(\Theta+\sigma)^{\prime}=\frac{(3\tilde \sigma C)^\prime}{C}+24\pi q_I B,
\label{ntex1}
\end{equation}
\begin{equation}
2(\Theta+\sigma)_{,\theta}=\frac{(3\tilde \sigma C)_{,\theta}}{C}+24\pi q_{II} Br,
\label{ntex2}
\end{equation}
which by virtue of (\ref{nthet}), become 
\begin{equation}
 \left(\frac{\dot B}{B}\right)^{\prime}=\frac{\dot B}{B}\frac{C^\prime}{C}-\frac{\dot C^\prime}{C}+8\pi q_I B=0,
\label{vf10}
\end{equation}
\begin{equation}
 \left(\frac{\dot B}{B}\right)_{,\theta}=\frac{\dot B}{B}\frac{C_{,\theta}}{C}-\frac{\dot C_{,\theta}}{C}+8\pi q_{I I}rB,
\label{vf11}
\end{equation}
where (\ref{vf4}) has been used.

Feeding back these two last equations into (\ref{vf7}, \ref{vf8}), and using (\ref{vf4}), we obtain
\begin{equation}
H_1=4\pi q_{II},\qquad H_2=-4\pi q_{I}.
\label{vf12}
\end{equation}

Next, combining (\ref{dB11}, \ref{48Y}) with (\ref{vf9}, \ref{vf12}) we obtain
\begin{equation}
(q_{II}BrC)^\prime-(q_IBC)_{,\theta}=0.
\label{vf13}
\end{equation}

 From (\ref{vf10}), (\ref{vf11}), using (\ref{vfn}) and (\ref{vf4}), we obtain
\begin{equation}
\frac{\dot T C^\prime-T\dot C^\prime}{\dot TC_{,\theta}-T\dot C_{,\theta}}=\tilde q(r,\theta)\equiv \frac{q(r,\theta)}{r},
\label{vfn3}
\end{equation}
whose first integral may be written as
\begin{equation}
T(t)R(r,\theta)=C^\prime-C_{,\theta}\tilde q(r,\theta),
\label{vfn4}
\end{equation}
where $R(r,\theta)$ is an arbitrary integration function.

The above equation, in turn, produces
\begin{equation}
C=T(t)\tilde C(r,\theta),
\label{vfn5}
\end{equation}
with
\begin{equation}
R(r,\theta)\equiv \tilde C^\prime-\tilde C_{,\theta}\tilde q(r,\theta).
\label{vfn6}
\end{equation}
Feeding back (\ref{vfn5}) into (\ref{vf10}) and (\ref{vf11}), it follows at once that $q_I=q_{II}=0$, which by virtue of (\ref{vf12}) implies that the magnetic part of the Weyl tensor vanishes. Furthermore, since the fluid is geodesic and non--dissipative it must be FRW, as shown in  \cite{2}.

We shall next consider the case where the vorticity is not assumed to vanish from the beginning.
\\
\section{The non-vanishing vorticity case}
In the previous section we have shown that, if the vanishing vorticity condition is assumed {\it ab initio}, then, the fluid configuration is just FRW. We shall now prove that this is the case even if such a condition is not imposed from the beginning. 

Thus,   let us now relax the vorticity--free condition. 

In this case we shall proceed as follows. Contracting (\ref{eim}) with vectors $\bold{K}$ and $\bold{L}$, we obtain respectively
\begin{eqnarray}
\Omega q_{II}&+&\frac{\kappa T_{,r}}{B \tau}+ \left \{\frac{1}{\tau}+\frac{1}{2}D_t \left[ln(\frac{\tau}{\kappa T^2})\right]-\frac{5}{6}\Theta \right\}q_I\nonumber \\ &-&\frac{q_I \sigma_I}{3}=0,
\label{eimk}
\end{eqnarray}
and 
\begin{eqnarray}
-\Omega q_{I}&+&\frac{\kappa L^\alpha T_{,\alpha}}{\tau}+ \left \{\frac{1}{\tau}+\frac{1}{2}D_t \left[\ln(\frac{\tau}{\kappa T^2})\right]-\frac{5}{6}\Theta \right\}q_{II}\nonumber \\ &-&\frac{q_{II}\sigma_{II}}{3}=0,
\label{eiml}
\end{eqnarray}
with $D_tf\equiv f_{,\beta}V^\beta$.

Let us now impose the regularity conditions at the center in  (\ref{eimk}, \ref{eiml}).

Close to the center, we may write
\begin{equation}
q_{II}\Omega=\sum_{n \geq N}q^{(n)}_{II}(t,\theta)\Omega^{(n)}(t,\theta) r^{n},
\label{sum1p}
\end{equation}
where $N>1$, and which implies because of (\ref{sum1})
\begin{equation}
q_{II}=\sum_{n \geq N-1}q^{(n)}_{II}(t,\theta)r^{n}.
\label{sum2p}
\end{equation}

Feeding back (\ref{sum1p})  into (\ref{eimk}), it follows at once that as $r\approx 0$,
\begin{equation}
q_{I}=\sum_{n \geq N}q^{(n)}_{I}(t,\theta)r^{n},
\label{sum3p}
\end{equation}
where the fact has beeen used that $ T_{,r}$ and $q_I$ have the same behaviour at the center, and of course, all physical and geometrical variables are regular  at $r\approx 0$.

But then, using (\ref{sum3p}) in (\ref{eiml}) it follows that 
\begin{equation}
q_{II}=\sum_{n \geq N+1}q^{(n)}_{II}(t,\theta)r^{n},
\label{sum4p}
\end{equation}
which of course is more restrictive than  (\ref{sum2p}).

We must then assume (\ref{sum4p}) instead of (\ref{sum2p}), in which  case (\ref{eimk}) produces
\begin{equation}
q_{I}=\sum_{n \geq N+2}q^{(n)}_{I}(t,\theta)r^{n},
\label{sum5p}
\end{equation}
implying because of (\ref{eiml})
\begin{equation}
q_{II}=\sum_{n \geq N+3}q^{(n)}_{II}(t,\theta)r^{n}.
\label{sum6p}
\end{equation}

Going through this cycle, once and again, as many times as desired, it is obvious that
\begin{equation}
q_{I}=\sum_{n \geq N}q^{(n)}_{I}(t,\theta)r^{n},\qquad \
q_{II}=\sum_{n \geq N}q^{(n)}_{II}(t,\theta)r^{n},
\label{sum7p}
\end{equation}
where now $N$ is an arbitrarily large number. In other words, not only $q_{I, II}$, but also all their  $r$-derivatives of any order, vanish at the center. This in turn implies that we can proceed exactly as in subsections II.B and II.C in \cite{2}, and prove that close to the center we also have $H^{(n)}_1=H^{(n)}_2={\cal E}^{(n)}_{I}={\cal E}^{(n)}_{II}=\sigma^{n}_1=\sigma^{n}_{II}=\Omega^{(n)}=0$ for any value of $n> 0$.

Then,  assuming that $q_{I, II}$ are of class $C^\omega$, we can analytically continue the zero value at the center to the whole configuration, i.e. we have a non--dissipative fluid. In this case as we know from \cite{2},  the system is a FRW.

\section{Conclusions}
At the light of the results obtained in \cite{2}, we started this research harboring the hope, that the inclusion of the dissipative term in the geodesic fluid distribution, would be enough to provide us with a  well behaved source of gravitational radiation. However, the final result indicates that the geodesic condition is  too restrictive, leading to the impossibility of ensuring regularity conditions at the center, in the presence of a dissipative flux (i.e. bringing us back to the case analyzed in \cite{2}).

But, could we guess such a result, before carrying  out the discussion presented here, only based on the geodesic condition? As we shall see now, the answer to this question, might be yes.

Indeed, in the $\Omega=0$ subcase, the above mentioned result might have been inferred  based on the link between vorticity and gravitational radiation \cite{vr}.

To see how this affirmative answer might have been guessed, in the second subcase (i.e. when the vorticity is not assumed to vanish from the beginning), let us recall the evolution equation for the vorticity scalar (Eq. B5 in \cite{1}, for  the geodesic case), which reads
\noindent 
\begin{equation}
\dot \Omega +\frac{1}{3}\left(2\Theta+\sigma_{I}+\sigma_{II}\right)\Omega=0.
\label{ecc4Y}
\end{equation}

Now,   let us  ignore, for a moment, the result obtained in this paper, about the fact that the geodesic condition leads to a FRW spacetime, implying the vanishing of vorticity, and let us focus on the equation above.

From (\ref{ecc4Y}) we see that the vanishing of vorticity at any given time, implies its vanishing at any other time afterwards, or, alternatively,  the presence of a non--vanishing vorticity in a geodesic fluid distribution, at any given time, implies that such a vorticity was there {\it always}, before that time. In other words, any possible non--vanishing vorticity in such  a system, would be a {\it primordial} one. Such a conclusion, of course, is independent on whether or not the system is dissipative. 

Now, the situation depicted above prevents the possibility to have an initially static system, which at some given moment departs from equilibrium and starts to evolve with a non vanishing vorticity, during a finite period of time. But then, and based on the link between vorticity and gravitational radiation, mentioned before, we might conclude, just from (\ref{ecc4Y}), that gravitational radiation from a physically meaningful system, i.e. one which radiates for a finite period of time, is not to be expected under the geodesic condition, even though the dissipation is not  excluded {\it ab initio}.

Of course, all these ``hand waving'' arguments, as sound as they may be, only provide the basis for a conjecture, and cannot replace the formal proof presented in sections III and IV.

The following comments are in order before ending:
\begin{enumerate}
\item In \cite{2}, it was shown that, only using the evolution equation for the vorticity, for the perfect (i.e. in thermodynamic equilibrium) and,  non--geodesic, fluid, the vorticity is also a {\it primordial} one.
Indeed, in this latter case, the evolution equation for the vorticity becomes
\begin{equation}
\Omega _{,\delta}V^\delta +\frac{1}{3}(2\Theta+\sigma _I+\sigma _{II}+V^\mu \Gamma_{,\mu})\Omega =0.\label{esc51KLb}
\end{equation}
where $\Gamma=\ln T$, and  $T$ denotes the  temperature.
Thus, even if the fluid is not geodesic, but is non--dissipative, the situation is the same as in the geodesic case, i.e. the vanishing of vorticity at any given time implies its vanishing for any time in the future.

  We then noticed that such a  result was in agreement  with earlier works indicating that vorticity is produced by dissipation \cite{Croco}--\cite{75}, which was absent in that discusssion. Here we do have a dissipative term, but only before enforcing the regularity conditions, and nevertheless the nature of the vorticity is of the same kind as that of the  dissipationless case. Of course, once regularity conditions are enforced, the resulting fluid is just FRW,  implying vorticity  and dissipation, vanish.
\item We must recall that the line element (\ref{metric2}), for which our results were obtained,  is not the most general, compatible with the axial symmetry. In particular, in the spherically symmetric limit, it describes only shear--free fluids. This is so because the shear, is now sourced by the magnetic parts of the Weyl tensor.

 Indeed, from (\ref{vf7}) and (\ref{vf8})
 we find that if the magnetic part of the Weyl tensor vanishes, then 

\begin{equation}
\tilde \sigma C=\psi(t),
\label{nu1}
\end{equation}
where $\psi(t)$ is an arbitrary integration function. The above equation implies, because of the regularity condition (\ref{nuev1}), that $\tilde \sigma=0$.
\item Models compatible with the geodesic condition do exist, if regularity conditions at the center are relaxed.
\item If regularity conditions at the center are maintained, then, physically acceptable models require the inclusion of, both, dissipative and anisotropic stresses terms, i.e. the geodesic condition must be abandoned. In this case, purely analytical methods are unlikely  to be sufficient to arrive at a full description of the source, and one has to resort to numerical methods.

 \end{enumerate}

\begin{acknowledgments}
L.H. thanks  the Departament de F\'isica at the  Universitat de les  Illes Balears, for financial support and hospitality. ADP  acknowledges hospitality of the
 Departament de F\'isica at the  Universitat de les  Illes Balears. J.O. acknowledges financial support from the Spanish
Ministry of Science and Innovation (Grant FIS2009-07238).
\end{acknowledgments}

\appendix* 
\section{Summary of equations for the geodesic dissipative case}
Below, we shall write only the equations  required for our discussion, and specialized for the geodesic and dissipative dust, from  the framework developed in \cite{1}. 

The equation (A7) in \cite{1}, which comes from the conservation law $T^\alpha _{\beta;\alpha}=0$, leads in our case to: 
\begin{equation}
h_{\alpha}^\beta
q_{\beta;\mu}V^\mu+
\left(\frac{4}{3}\Theta h_{\alpha \beta}+\sigma_{\alpha
\beta}+\Omega_{\alpha\beta}\right )q^\beta=0.\label{ec2}
\end{equation}
which is the ``generalized Euler'' equation.  

\noindent From {\bf B.10}, {\bf B.12}, {\bf B.13}, in \cite{1}, which are some of the scalar equations resulting  from the projections of the Bianchi identities with the tetrad vectors, we obtain respectively
\begin{widetext}
\begin{eqnarray}
\frac{1}{3}\dot{\mathcal E}_{I}+\frac{4\pi}{3}\mu\sigma_{I}-\Omega \mathcal E_{KL}
+\frac{\mathcal E_{I}}{9}\left(3\Theta+\sigma_{II}-\sigma_{I}\right)+\frac{\mathcal E_{II}}{9}\left(2\sigma_{II}+\sigma_{I}\right)\nonumber\\
-\frac{1}{\sqrt{B^2r^2+\tilde G^2}}\left(H_{1,\theta}+H_1\frac{C_{,\theta}}{C}\right)
-\frac{H_2}{B}\left[
\frac{C^{\prime}}{C}-\frac{2(Br)(Br)^{\prime}+\tilde G \tilde G^\prime}{2\left(B^2r^2+\tilde G^2\right)}\right]\nonumber
\\
=-\frac{4\pi}{B}q^\prime _{I}-\frac{4\pi q_{II}}{\sqrt{B^2r^2+\tilde G^2}}\left(\tilde G\frac{\dot B}{B}+\frac{B_{,\theta}}{B}\right),
\label{dB10Y}
\end{eqnarray}

\noindent 
\begin{eqnarray}
\frac{1}{3}\dot{\mathcal E}_{II}+\frac{4\pi}{3}\mu\sigma_{II}+\Omega \mathcal E_{KL}
+\frac{\mathcal E_{II}}{9}\left(3\Theta+\sigma_{I}-\sigma_{II}\right)+\frac{\mathcal E_{I}}{9}\left(2\sigma_{I}+\sigma_{II}\right)\nonumber\\
+\frac{1}{B}\left(H_{2}^{\prime}+H_{2}\frac{C^{\prime}}{C}\right)
+\frac{H_{1}}{\sqrt{B^2r^2+\tilde G^2}}\left[\frac{C_{,\theta}}{C}-\frac{B_{,\theta}}{B}-\tilde G\left(\frac{\dot B}{B}-\frac{\dot C}{C}\right)\right]\nonumber
\\
=-2\pi \frac{q_I}{B}\frac{(B^2r^2+\tilde G^2)^\prime}{B^2r^2+\tilde G^2}-\frac{4\pi(\tilde G \dot q_{II}+q_{II,\theta})}{\sqrt{B^2r^2+\tilde G^2}},
\label{dB12}
\end{eqnarray}

\noindent 
\begin{eqnarray}
-\frac{1}{3}\left(\mathcal E_{I}+\mathcal E_{II}\right)^{.}-\frac{1}{3}(\mathcal E_{I}+\mathcal E_{II})\Theta-\frac{4\pi}{3}\mu(\sigma_{I}+\sigma_{II})
-\frac{\mathcal E_{I}}{9}\left(2\sigma_{II}+\sigma_{I}\right)-\frac{\mathcal E_{II}}{9}\left(2\sigma_{I}+\sigma_{II}\right)\nonumber\\
+\frac{1}{\sqrt{B^2r^2+\tilde G^2}}\left(H_{1,\theta}+H_{1}\frac{B_{,\theta}}{B}\right)
-\frac{1}{B}\left\{H_{2}^{\prime}+H_{2}\left[\frac{(B^2r^2+\tilde G^2 )^{\prime}}{2(B^2r^2+\tilde G^2)}\right]\right\}\nonumber
\\
=-\frac{4\pi q_I}{B}\frac{C^\prime}{C}-\frac{4\pi q_{II}}{\sqrt{B^2r^2+\tilde G^2}}\left(\tilde G\frac{\dot C}{C}+\frac{C_{,\theta}}{C}\right).\nonumber\\
\label{dB13}
\end{eqnarray}

A combination of (\ref{dB12}) and (\ref{dB13}) produces
\begin{eqnarray}
-\frac{1}{3} \dot{\mathcal E}_{I}-\frac{4\pi}{3}\mu\sigma_{I}+\Omega \mathcal E_{KL}
-\frac{\mathcal E_{I}}{9}(3\Theta-\sigma_{I}+\sigma_{II})-\frac{\mathcal E_{II}}{9}(\sigma_{I}+2\sigma_{II}) \nonumber\\
+\frac{1}{\sqrt{B^2r^2+\tilde G^2}}\left \{H_{1,\theta}+H_1\left[\frac{C_{,\theta}}{C}-\tilde G\left(\frac{\dot B}{B}-\frac{\dot C}{C}\right)\right]\right \}
+\frac{H_2}{B}\left [\frac{C^\prime}{C}-\frac{(B^2r^2+\tilde G^2)^\prime}{2(B^2r^2+\tilde G^2)}\right ]\nonumber \\
=- \frac{4\pi}{B}q_I\left [\frac{C^\prime}{C}+\frac{(B^2r^2+\tilde G^2)^\prime}{2(B^2r^2+\tilde G^2)}\right ]-\frac{4\pi}{\sqrt{B^2r^2+\tilde G^2}}\left [\tilde G \dot q_{II}+q_{II,\theta}+ q_{II}\left(\tilde G\frac{\dot C}{C}+\frac{C_{,\theta}}{C}\right)\right ],
\label{44Y}
\end{eqnarray}
whereas from  (\ref{dB10Y}) with (\ref{44Y}) we obtain
\begin{eqnarray}
\frac{\tilde G}{\sqrt{B^2r^2+\tilde G^2}}\left[\frac{H_2 \tilde G^\prime}{2 B \sqrt{B^2r^2+\tilde G^2}}+H_1 \left(\frac{\dot B}{B}-\frac{\dot C}{C}\right)\right]\nonumber
\\
=\frac{4\pi}{B}\left\{q^\prime _I+q_I\left[\frac{C^\prime}{C}+ \frac{(B^2r^2+\tilde G^2)^\prime}{2(B^2r^2+\tilde G^2)}\right]\right \}\nonumber
\\
+\frac{4\pi }{\sqrt{B^2r^2+\tilde G^2}}\left\{q_{II}\left[\tilde G\left(\frac{\dot C}{C}+\frac{\dot B}{B}\right)+\frac{C_{,\theta}}{C}+\frac{B_{,\theta}}{B}\right]
+\tilde G \dot q_{II}+q_{II,\theta}\right\}.
\label{45Y}
\end{eqnarray}

Next,  from {\bf B.6},  {\bf B.7},  {\bf B.8}, and  {\bf B.9} of \cite{1}, which are obtained by contracting the Ricci identities with the tetrad vectors, we obtain 
\begin{eqnarray}
-\frac{1}{\sqrt{B^2r^2+\tilde G^2}}\left [ \Omega _{,\theta}+\tilde G \dot \Omega+ \Omega\left(\frac{\tilde G\dot C}{C}+\frac{C_{,\theta}}{C}\right )\right]
+\frac{1}{3B}\left\{2 \Theta^\prime-\sigma_{I}^\prime
-\sigma_{I}\left[\frac{2C^{\prime}}{C}+\frac{\left(B^2r^2+\tilde G^2\right)^{\prime}}{2\left(B^2r^2+\tilde G^2\right)}\right]\right.\nonumber\\
\left.-\sigma_{II}\left[\frac{C^{\prime}}{C}-\frac{\left(B^2r^2+\tilde G^2\right)^{\prime}}{2\left(B^2r^2+\tilde G^2\right)}\right]\right\}=8\pi q_I,\nonumber\\
\label{ecc5Y}
\end{eqnarray}

\begin{eqnarray}
\frac{1}{B}\left(\Omega^{\prime}+\Omega \frac{C^{\prime}}{C}\right)
+\frac{1}{3\sqrt{B^2r^2+\tilde G^2}}\left\{\left(2\Theta-\sigma_{II}\right)_{,\theta}+\tilde G \left(2\Theta-\sigma_{II}\right)^{.}\right.\nonumber\\
\left.+\sigma_{I}\left[\frac{B_{\theta}}{B}-\frac{C_{\theta}}{C}+\tilde G\left(\frac{\dot B}{B}-\frac{\dot C}{C}\right)\right]-\sigma_{II}\left[\frac{B_{\theta}}{B}+\frac{2C_{\theta}}{C}+\tilde G\left(\frac{\dot B}{B}+\frac{2\dot C}{C}\right)\right]\right\}=8\pi q_{II},\nonumber\\
\label{ecc6Y}
\end{eqnarray}

\begin{eqnarray}
H_1=-\frac{1}{2B}\left\{\Omega^{\prime}-\Omega\left[\frac{C^{\prime}}{C}-\frac{\tilde G \tilde G^{\prime}}{2(B^2r^2+\tilde G^2)}\right]\right\}
-\frac{1}{6\sqrt{B^2r^2+\tilde G^2}}\left\{\left(2\sigma_{I}+\sigma_{II}\right)_{,\theta}\right.\nonumber\\
\left.+\sigma_{I}\left[\frac{B_{,\theta}}{B}+\frac{C_{,\theta}}{C}-\tilde G(\frac{\dot B}{B}-\frac{\dot C}{C})\right]-\sigma_{II}\left[\frac{B_{,\theta}}{B}-\frac{2C_{,\theta}}{C}+\tilde G(\frac{2\dot B}{B}-\frac{2\dot C}{C})\right]\right\},
\label{ecc8Y}
\end{eqnarray}

\begin{eqnarray}
H_2=+\frac{1}{6B}\left\{\left(\sigma_{I}+2\sigma_{II}\right)^{\prime}+\sigma_{I}\left[\frac{2C^\prime}{C}-\frac{(Br)(Br)^{\prime}}{B^2r^2+\tilde G^2}\right]
+\sigma_{II}\left[\frac{C^\prime}{C}+\frac{2(Br)(Br)^{\prime}+3\tilde G \tilde G^\prime}{2\left(B^2r^2+\tilde G^2\right)}\right]\right\}\nonumber\\
-\frac{1}{2\sqrt{B^2r^2+\tilde G^2}}\left\{\Omega_{,\theta}-\Omega \left[\frac{C_{,\theta}}{C}+\tilde G\left(\frac{\dot B}{B}+\frac{\dot C}{C}\right)\right]\right\},\nonumber
\\
\label{ecc7Y}
\end{eqnarray}

and  from {\bf B.11} and  {\bf B.16} of \cite{1}, which are scalar equations, also obtained from the Bianchi identities
\begin{eqnarray}
2\dot{\mathcal E}_{KL}+\mathcal E_{KL}\left(2\Theta-\sigma_{I}-\sigma_{II}\right)+\frac{\Omega}{3}\left(\mathcal E_{I}-\mathcal E_{II}\right)
+\frac{1}{B}\left\{H_{1}^{\prime}+H_{1}\left[\frac{2C^{\prime}}{C}-\frac{2(Br)(Br)^{\prime}+\tilde G \tilde G^{\prime}}{2\left(B^2r^2+\tilde G^2\right)}\right]\right\}\nonumber\\
-\frac{1}{\sqrt{B^2r^2+\tilde G^2}}\left\{H_{2,\theta}+H_{2}\left[\frac{2C_{,\theta}}{C}-\frac{B_{,\theta}}{B}
-\tilde G\left(\frac{\dot B}{B}-\frac{\dot C}{C}\right)\right]\right\}\nonumber
\\
= \frac{2\pi}{\sqrt{B^2r^2+\tilde G^2}}\left [q_I\left(\tilde G \frac{\dot B}{B}+\frac{B_{,\theta}}{B}\right)-\tilde G \dot q_I-q_{I,\theta}\right ]+\frac{2\pi}{B}\left [-q^\prime _{II}+q_{II}\frac{(B^2r^2+\tilde G^2)^\prime}{2(B^2r^2+\tilde G^2)}\right ],
\label{dB11}
\end{eqnarray}

\begin{eqnarray}
 \frac{1}{3}\mathcal E_{KL}\left(\sigma_{II}-\sigma_{I}\right)-\frac{1}{B}\left\{H_1^\prime + H_1\left[\frac{2C^\prime}{C}
+\frac{(B^2r^2+\tilde G^2)^\prime}{2(B^2r^2+\tilde G^2)}\right]\right\}\nonumber\\
-\frac{1}{\sqrt{B^2r^2+\tilde G^2}}\left\{H_{2,\theta}+\tilde G \dot H_{2}+H_2 \left[\frac{B_{,\theta}}{B}+\frac{2C_{,\theta}}{C}+\tilde G\left(\frac{\dot B}{B}+\frac{2 \dot C}{C}\right)\right]\right\}
\nonumber\\
=\left[8\pi\mu-(\mathcal{E}_I+\mathcal E_{II})\right]\Omega-\frac{4\pi}{\sqrt{B^2r^2+\tilde G^2}}\left (q_{I,\theta}+q_I\frac{B_{,\theta}}{B}\right )+\frac{4\pi}{B} \left [q^\prime _{II}+q_{II}\frac{(B^2r^2+\tilde G^2)^\prime}{2(B^2r^2+\tilde G^2)}\right ].\nonumber\\
\label{48Y}
\end{eqnarray}
Finally, the transport equation for the geodesic case, (Eq.(57) in \cite{1}) reads:
\begin{equation}
\tau h^\mu_\nu q^\nu _{;\beta}V^\beta +q^\mu=-\kappa
h^{\mu\nu}T_{,\nu}-\frac{1}{2}\kappa T^2\left
(\frac{\tau V^\alpha}{\kappa T^2}\right )_{;\alpha}q^\mu,\label{qT}
\end{equation}

then, from a combination of (\ref{ec2}) and (\ref{qT})  (Eq.(60) in \cite{1}), we obtain in our case.

\begin{eqnarray}
-(\sigma _{\alpha \beta}+\Omega _{\alpha \beta})q^\beta
+\frac{\kappa}{\tau}\nabla _\alpha T+\left \{\frac{1}{\tau}+\frac{1}{2}D_t \left[ln(\frac{\tau}{\kappa T^2})\right]-\frac{5}{6}\Theta \right\}q_\alpha=0.
\label{eim}
\end{eqnarray}
\end{widetext}

\end{document}